\documentclass[%
 reprint,
superscriptaddress,
 amsmath,amssymb,
 aps,
pra,
]{revtex4-2}

\usepackage{graphicx}
\usepackage{dcolumn}
\usepackage{bm}
\usepackage{mathtools}
\usepackage{float}
\usepackage{siunitx}

\begin{document}

\preprint{APS/123-QED}

\title{Investigating the hyperfine systematic error and relative phase in low spin-polarization alkali FID magnetometers}

\author{D. P. Hewatt}\email{Dawson.Hewatt@colorado.edu}
\affiliation{Department of Physics, University of Colorado, Boulder, Colorado 80309, USA}
\affiliation{Paul M. Rady Department of Mechanical Engineering, University of Colorado, Boulder, Colorado 80309, USA}
\affiliation{JILA, National Institute of Standards and Technology and University of Colorado, Boulder, Colorado 80309, USA}
\author{M. Ellmeier}
\affiliation{Paul M. Rady Department of Mechanical Engineering, University of Colorado, Boulder, Colorado 80309, USA}
\author{C. Kiehl}
\affiliation{Department of Physics, University of Colorado, Boulder, Colorado 80309, USA}
\affiliation{JILA, National Institute of Standards and Technology and University of Colorado, Boulder, Colorado 80309, USA}
\author{T. S. Menon}
\affiliation{Department of Physics, University of Colorado, Boulder, Colorado 80309, USA}
\affiliation{JILA, National Institute of Standards and Technology and University of Colorado, Boulder, Colorado 80309, USA}
\author{J. W. Pollock}
\affiliation{Paul M. Rady Department of Mechanical Engineering, University of Colorado, Boulder, Colorado 80309, USA}
\author{C. A. Regal}
\affiliation{Department of Physics, University of Colorado, Boulder, Colorado 80309, USA}
\affiliation{JILA, National Institute of Standards and Technology and University of Colorado, Boulder, Colorado 80309, USA}
\author{S. Knappe}
\affiliation{Paul M. Rady Department of Mechanical Engineering, University of Colorado, Boulder, Colorado 80309, USA}
\affiliation{FieldLine Medical, Boulder Colorado 80301, USA}
\affiliation{FieldLine Industries, Boulder Colorado 80301, USA}

\date{\today}
\begin{abstract}
Alkali-metal optically pumped magnetometers that utilize vapor cells containing high buffer-gas pressure and operate at elevated temperatures are prone to inaccuracies arising from the overlap of the average $F=I+1/2$ and $F=I-1/2$ ground-state Zeeman resonances. We employ density-matrix simulations and experiments to investigate how this hyperfine systematic error varies with spin polarization in a $^{87}$Rb free-induction-decay (FID) magnetometer. At low spin polarizations $P~\leq~0.5$, this effect causes single-frequency magnetic-field extraction techniques to exhibit inaccuracies up to approximately 3.5 nT. Density-matrix simulations reveal that this bias can be traced to the relative amplitude and phase between the $F=I~{\pm}~1/2$ hyperfine ground-state manifolds in the FID spin precession signal. We show that this systematic error can be mitigated using either a double-frequency fitting model that accounts for the relative amplitude and phase or synchronous-pulse pumping that minimizes the $F=1$ contribution to the FID signal. Theoretical simulations predict accuracies within 0.5~nT for both techniques across a wide range of spin polarizations, suggesting a sevenfold enhancement over single-frequency extraction methods. Our experiments validate this, showcasing a variation in the extracted field below 1~nT, a 3.5-fold improvement compared to single-frequency extraction methods. Furthermore, the mitigation techniques demonstrate agreement of the extracted magnetic field within 1.5~nT.
\end{abstract}

\maketitle

\section{\label{sec:intro}Introduction}

\begin{figure*}[hbt!]
\includegraphics[width=\linewidth]{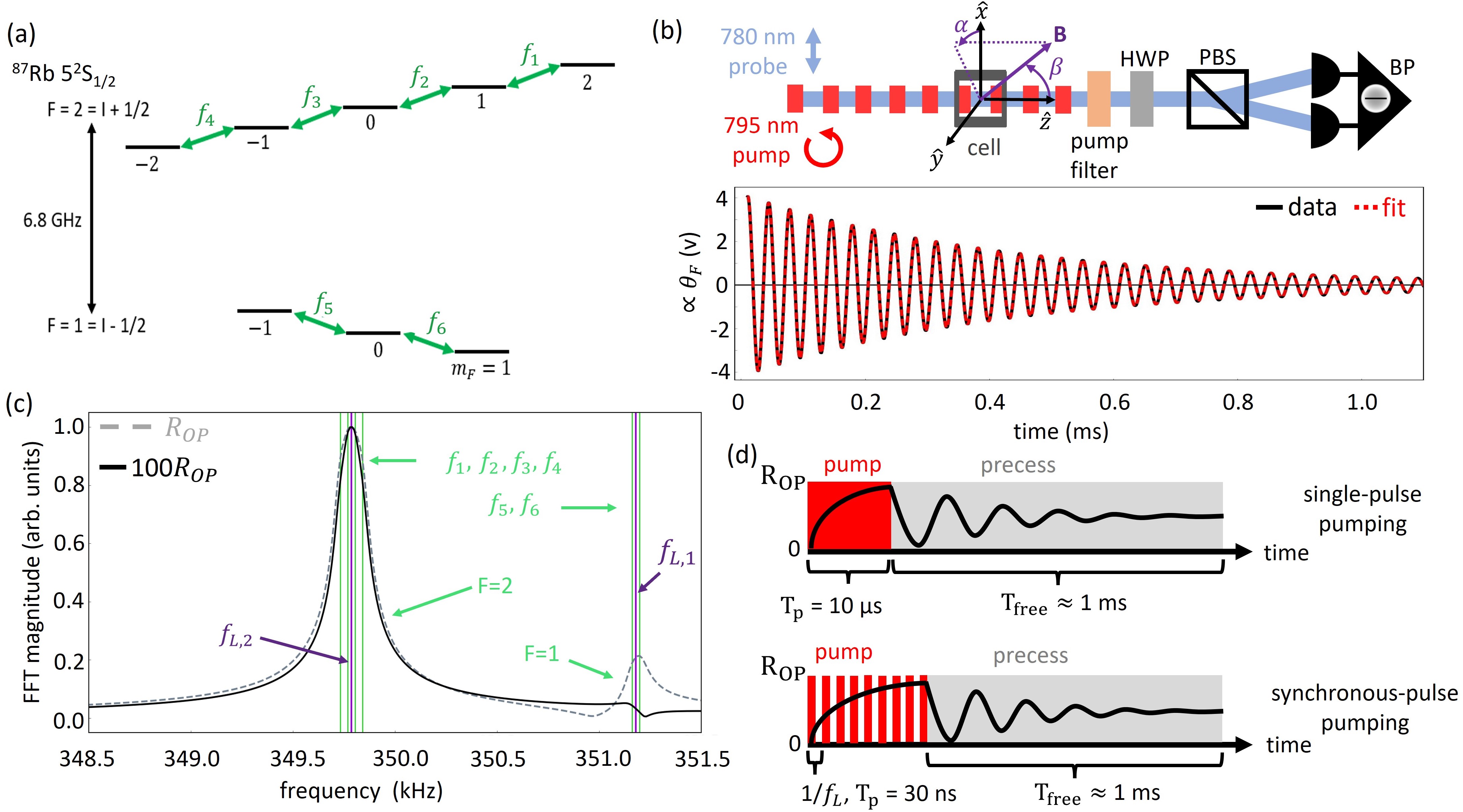}
\caption{\label{fig:FIDBackgorund} (a) Level diagram for the $^{87}$Rb ground state. Free-induction-decay magnetometers measure the Zeeman resonance frequencies $f_{i}$ shown in green. (b) Magnetometer apparatus and example FID signal. Here, HWP is the half-wave plate; PBS is a Wollaston polarizing beam splitter, and BP is a balanced polarimeter. The magnetic field \textbf{B} is specified by a magnitude B azimuthal angle $\alpha$, and polar angle $\beta$. The bottom graph shows an FID signal (black) and single-frequency fit (red). (c) Normalized Fourier transforms of simulated FID signals, with black solid and gray dashed lines representing high (100$R_{\text{OP}}$) and low ($R_{\text{OP}}$) pumping rates, respectively. These simulations neglect spin exchange so that the $F=I~{\pm}~1/2$ resonances could be resolved. The green vertical lines are the individual Zeeman transitions at \SI{50}{\micro\tesla}, while the purple vertical lines are the average Larmor frequencies for the $F=I~{\pm}~1/2$ hyperfine manifolds. (d) FID pumping schemes for single-pulse pumping (top) and synchronous-pulse pumping (bottom). Single-pulse pumping uses one pump pulse with a pumping rate $R_{\text{OP}}$ and pulse length $T_\text{p}$. Synchronous-pulse pumping modulates the amplitude of the pump laser at the average Larmor frequency $f_L$. Note that the time axes for these graphs are not to scale. After pumping, the atoms precess in the dark for a time $T_\text{free}$.}
\end{figure*}

Scalar optically pumped magnetometers (OPMs), known for their high sensitivity, robustness, and ability to be miniaturized, have found widespread applications in fields such as fundamental physics \cite{abel2020measurement, abel2017search, groeger2006laser}, space magnetometry \cite{korth2016miniature, bennett2021precision}, and biomedical imaging \cite{tierney2019optically, boto2017new, bison2003dynamical, sander2012magnetoencephalography}. The accuracy of alkali-metal OPMs is limited by frequency shifts of the unresolved Zeeman sublevels, caused by effects such as the nonlinear Zeeman (NLZ) shift (Fig.~\ref{fig:FIDBackgorund}(a)). Microfabricated vapor cells often contain buffer gas to reduce depolarizing wall collisions and operate at elevated temperatures (80~$^{\circ}$C\text{--}120~$^{\circ}$C) to increase atomic density, enhancing sensitivity. The addition of buffer gas broadens and shifts the optical resonances hindering hyperfine-selective pumping, while elevated temperatures increase spin-exchange collisions, broadening the magnetic resonance. Since the $F=I~{\pm}~1/2$ manifolds have different hyperfine Land\'e $g$-factors, there is a frequency difference between the average Zeeman resonances of the ground-state hyperfine manifolds \cite{lee2021heading, Steck87Rb}. In regimes of high buffer gas and elevated temperatures, the average Zeeman resonances of the $F=I~{\pm}~1/2$ hyperfine manifolds become spectrally unresolved, introducing an additional source of inaccuracy we call the hyperfine systematic error.

Techniques like analytical correction terms \cite{lee2021heading}, double-pass configurations \cite{rosenzweig2023heading}, light-polarization modulation \cite{oelsner2019sources}, double-modulated synchronous pumping \cite{seltzer2007synchronous}, and spin-locking \cite{bao2018suppression} can mitigate NLZ effects. While these approaches effectively minimize NLZ shifts, they do not account for the hyperfine systematic error, limiting their accuracy in regimes of low initial atomic spin polarization $P\leq0.5$, where this error becomes non-negligible. Techniques utilizing higher-order polarization modes are inherently immune to the NLZ effect and hyperfine systematic error \cite{zhang2023heading, acosta2008production}. However, the transition amplitudes of these higher-order modes are significantly suppressed in vapor cells with high buffer-gas pressure, limiting the sensitivity \cite{nagel1999precision, rushton2023alignment}. To address the hyperfine systematic error, Lee \textit{et al.} proposed averaging precession signals between two orthogonal probe beams, but this approach requires a second probe laser, increasing sensor complexity \cite{lee2021heading}.

Our analysis reveals that unresolved Zeeman resonances between the $F=I~{\pm}~1/2$ manifolds cause techniques that use a single frequency to extract the magnetic field to produce inaccuracies exceeding 1 nT for atomic spin polarizations below 0.5 (Fig.~\ref{fig:SimAcc}(a)) \cite{lee2021heading,hunter2023optical}. Low-power consumption systems, such as compact sensors, often lack access to the high-powered lasers required for generating high spin polarizations in optically dense media used in microfabricated cells. Therefore, these magnetometers are susceptible to these accuracy constraints. This systematic error is a barrier to applications like navigation, fundamental physics, and space magnetometry, where the demand for accuracies less than or equal to 1~nT is commonplace \cite{canciani2016absolute, grujic2015sensitive, fratter2016swarm}.

There are many OPM detection methods including Bell-Bloom \cite{bell1961optically, gerginov2017pulsed}, $M_x$ and $M_z$ \cite{groeger2006laser, schwindt2007chip}, coherent population trapping (CPT) \cite{ellmeier2023frequency, stahler2002coherent, schwindt2004chip}, Rabi \cite{thiele2018self, kiehl2023coherence, kiehl2023correcting}, spin-exchange-relaxation-free regime \cite{allred2002high, ledbetter2008spin}, and free-induction-decay regime (FID) \cite{hunter2018free, afach2015highly, fabricant2023build}. Hyperfine measurement approaches such as CPT and microwave interrogation techniques can overcome these systematic inaccuracies because they can resolve selected transitions \cite{pollinger2018coupled, liang2014simultaneously, kiehl2023correcting}. However, these hyperfine techniques exhibit limited sensitivity when compared to methods that measure Zeeman resonances. Spin-exchange-relaxation-free OPMs are well known for displaying sub-fT/$(\text{Hz})^{1/2}$ sensitivities, but can only operate at near-zero fields \cite{kominis2003subfemtotesla, dang2010ultrahigh, sheng2013subfemtotesla}. In contrast, FID magnetometers can operate at geomagnetic fields (approximately \SI{50}{\micro\tesla}) and have demonstrated  fT/$(\text{Hz})^{1/2}$-level sensitivities \cite{gerginov2020scalar, grujic2015sensitive, lee2021heading, hunter2023optical}. Unlike driven magnetometers, FID sensors do not require prior knowledge of the magnetic field to operate. This can improve the magnetic field slew rate by eliminating the need for an additional control loop \cite{hunter2018free}. Additionally, the pulsed operation used in FID magnetometers minimizes light shifts by off-resonantly probing the atomic state.

In an FID measurement scheme, the atoms are optically pumped, and then freely precess in the dark, sampling the six Zeeman splittings $f_{i}$ shown in Fig.~\ref{fig:FIDBackgorund}(a). Although there are various ways to implement an FID magnetometer, a schematic of the apparatus used in this study is shown in Fig.~\ref{fig:FIDBackgorund}(b). The magnetometer consists of a $3\times3\times2$~mm$^3$ micro-fabricated vapor cell containing enriched $^{87}$Rb and heated to roughly $108~^{\circ}\text{C}$ via resistive heaters. Approximately 1000~Torr of N$_2$ buffer gas is added to the cell to limit wall collisions. Circularly polarized laser light, resonant with the $^{87}$Rb $D_1$ transition, pumps the atoms into one of the stretched states $|F=2$,~$m_F=\pm2\rangle$. Free-induction-decay magnetometers commonly employ one of two primary pumping methods to create an initial spin polarization: single-pulse pumping with one uninterrupted pulse of pulse length $T_\text{p}$ or synchronous-pulse pumping with many short pulses, amplitude modulated (or frequency modulated) at the Larmor frequency (Fig.~\ref{fig:FIDBackgorund}(d)) \cite{hunter2018free}. A balanced polarimeter measures the precession of the atomic spin ensemble by detecting the Faraday rotation of a far-detuned, linearly polarized probe beam (Fig.~\ref{fig:FIDBackgorund}(b)). Here we use a collinear pump-probe geometry. The average Larmor frequency $f_L$ of the spin precession is extracted to determine the magnetic field.

As depicted in Fig.~\ref{fig:FIDBackgorund}(c), optical pumping affects the relative amplitudes of the various frequency components within FID signals and consequently the accuracy of the magnetometer. Higher optical pumping rates $R_{\text{OP}}$ generate larger initial spin polarizations. As the spin polarization decreases, the amplitude of the F~=~I~$-$~1/2 frequency component increases, shifting the average Larmor frequency. Furthermore, we also find that the relative phase between the $F=I~{\pm}~1/2$ Zeeman resonances depends on optical pumping and can affect the extracted field. 

In this paper we use simulation and experiment to investigate how the hyperfine systematic error affects the accuracy of magnetic-field extraction as a function of initial spin polarization in a $^{87}$Rb FID magnetometer. At low spin polarizations $P\leq0.5$, we demonstrate that implementing a double-frequency fitting model more accurately extracts the magnetic field. Conversely, synchronous-pulse pumping can enhance the initial spin polarization and consequently the accuracy of single-frequency extraction techniques. Theoretically, we show that both double-frequency fitting and synchronous-pulse pumping result in an enhancement of the FID accuracy by a factor of 7 compared to single-frequency fitting of FID signals produced by single-pulse pumping. Experimentally, we observe agreement within 1.5 nT between both the double-frequency and synchronous-pulse pumping techniques.

\section{\label{sec:TheoModel}Theoretical Model}

We describe the ground-state dynamics of this system using an $8\times8$ density matrix, $\rho$, in the total atomic spin basis basis $|F$, $m_F \rangle_z$, where the $z$ indicates that the basis is chosen in the $\hat{z}$ direction of the laboratory frame. The Hamiltonian that characterizes the hyperfine and Zeeman interactions of the $^{87}$Rb atom is given by
\begin{equation}
\label{eq:Hamiltonian}
 \begin{aligned}
H=&R_z(\alpha)R_y(\beta)\big\{\mathcal{M}\big[A_\text{hfs}\mathbf{I}\cdot\mathbf{S}+\mu_B(g_sS_z+g_iI_z)B\big]\mathcal{M}^{\dagger}\\&-I_2{\Delta}E\big\}R_y(\beta)^{\dagger}R_z(\alpha)^{\dagger},
 \end{aligned}
\end{equation}
where A$_\text{hfs}$ is the hyperfine splitting constant, \textbf{I} is the nuclear spin, \textbf{S} is the electron spin, B is the dc magnetic field strength, $\mu_B$ is the Bohr magneton, $g_s$ and $g_i$ are the electron and nuclear $g$-factors, respectively, as defined by \cite{Steck87Rb}, $I_2$ is the $F=2$ identity operator, and ${\Delta}E=h(6.834~\text{GHz})$ is the hyperfine splitting energy. Because the Hamiltonian is not generally diagonal in the $|F$, $m_F \rangle_z$ basis unless the magnetic field is in the $\hat{z}$ direction, we define $\mathcal{M}$ as the operator that diagonalizes the hyperfine and Zeeman terms \cite{kiehl2023correcting}. This serves to maintain the NLZ effects during the rotating-wave approximation (RWA). The first term in Eq. (\ref{eq:Hamiltonian}) describes the hyperfine splitting of the $^{87}$Rb atoms and the second term describes the Zeeman interaction between the atoms and the magnetic field. The third term in this Hamiltonian eliminates the high-energy scale caused by the hyperfine splitting, thereby enhancing computational efficiency. We rotate the system into the direction of the magnetic field by applying the rotation operators $R_z(\alpha)$~=~$e^{-iF_z\alpha}$ and $R_y(\beta)$~=~$e^{-iF_y\beta}$ (Fig. \ref{fig:FIDBackgorund}(b)). In $^{87}$Rb, the Zeeman effect causes splitting of the $m_F$ levels with a gyromagnetic ratio of approximately 7 Hz/nT. The energy eigenvalues of the $F=I~{\pm}~1/2$ sublevels are given by the Breit-Rabi formula \cite{breit1931measurement}
\begin{equation}
\label{eq:BR}
E_{m_F}=\frac{{\Delta}E}{2}\left(\frac{-1}{2I+1}+\frac{2g_{I}m_{F}x}{g_s-g_I}{\pm}\sqrt{x^2+\frac{4xm_F}{2I+1}+1}\right),
\end{equation}
where $x=\mu_B(g_s-g_I)B/{\Delta}E$ and $I=3/2$. The FID signal contains six frequency components given by the difference between adjacent Zeeman sublevels (Fig.~\ref{fig:FIDBackgorund}(a)).

The spin dynamics are approximated by the density matrix equation 
\begin{equation}
\label{eq:DenMat}
 \begin{aligned}
     \frac{d\rho}{dt}=&\frac{[H,\rho]}{i\hbar}+R_\text{SE}[\varphi(1+4\langle\mathbf{S}\rangle\cdot\mathbf{S})-\rho]+R_\text{SD}(\varphi-\rho)\\
     & +D\nabla^2\rho+R_\text{OP}[\varphi(1+2\mathbf{s}\cdot\mathbf{S})-\rho],
 \end{aligned}
\end{equation}
where $\varphi$~=~$\rho$/4~+~$\textbf{S}\cdot\rho\textbf{S}$ is the purely nuclear part of the density matrix \cite{appelt1998theory}. The first term is the time evolution of the Hamiltonian [Eq. (\ref{eq:Hamiltonian})]. The second term describes the effect of spin-exchange collisions between alkali-metal atoms at a rate $R_\text{SE}$. The third term expresses the effect of spin-destruction collisions between Rb with itself and N$_2$ at a rate of $R_\text{SD}$, where $R_\text{SD}$~=~$R^\text{Rb-Rb}_\text{SD}$~+~$R^\text{Rb-N$_2$}_\text{SD}$. The rates of these mechanisms are given by the definition, $R_\text{SE (SD)}~{\equiv}~n_\text{Rb (N$_2$)}\sigma_\text{SE (SD)}\textit{$v_{r}$}$ \cite{allred2002high}. Here $n_\text{Rb (N$_2$)}$ is the number density of the alkali-metal atoms and buffer-gas atoms, respectively, $\sigma_\text{SE (SD)}$ is the cross section of the collision, and $\textit{$v_{r}$}$ is the relative velocity of the atoms given by $\textit{$v_{r}$}=\sqrt{8k_{B}T/{\pi}M}$, where $k_B$ is the Boltzmann constant, $T$ is the temperature, and $M$ is the reduced mass of the colliding pair. The fourth term accounts for depolarizing wall collisions. We assume that the lowest spin-diffusion decay mode dominates the wall collision term such that we can make the approximation $D\nabla^2\rho~{\rightarrow}~R_\text{W}(\rho_{e}-\rho)$ where the wall-collision rate is $R_\text{W}$~=~$D\pi^2$/($\textit{$l_{x}^2$}$~+~$\textit{$l_{y}^2$}$~+~$\textit{$l_{z}^2$}$), $D$ is the diffusion constant, $l_i$ is the cell length along the \textit{i}th axis, and $\rho_{e}$ is the equilibrium density matrix containing equal populations in all states \cite{Jau2005new}. The last term describes optical pumping along a single axis in the limit of high buffer-gas pressure. This term is characterized by a pumping rate $R_{\text{OP}}$ and a mean photon spin \textbf{s}~=~s$\hat{\textit{z}}\in(-1,1)$ \cite{appelt1998theory}. Spin-exchange collisions are the dominant source of decoherence. Table \ref{tab:DecRates} shows the decoherence rates used in this paper. Since we applied the RWA to the Hamiltonian, we must also apply the RWA to the decoherence and pumping terms by numerically averaging such that the high-frequency counterrotating terms are eliminated \cite{kiehl2023coherence}.

\renewcommand{\arraystretch}{1.4}
\begin{table}[hbt!]
\begin{ruledtabular}
\begin{tabular}{c|c|c}
\textrm{Collision}&
\textrm{Cross section}&
\textrm{Rate~(s$^{-1}$)}\\
\colrule
(Rb-Rb)$_\text{SE}$ & $\sigma_\text{SE}$~=~1.90$\times10^{-18}~\text{m}^2$ & $R_\text{SE}$ = 6600 \\
(Rb-Rb)$_\text{SD}$ & $\sigma_\text{SD}^\text{Rb-Rb}$~=~1.77$\times10^{-21}~\text{m}^2$ & $R_\text{SD}^\text{Rb-Rb}$ = 6\\
(Rb-N$_2$)$_\text{SD}$ & $\sigma_\text{SD}^\text{Rb-N$_2$}$~=~1.44$\times10^{-26}~\text{m}^2$ & $R_\text{SD}^\text{Rb-N$_2$}$ = 226\\
Wall & $D$ = 1.69$\times10^{-5}~\text{m}^2s^{-1}$ & $R_\text{W}$ = 79
\end{tabular}
\end{ruledtabular}
\caption{\label{tab:DecRates} Decoherence rates for various collisions used in this paper. The cross sections are taken from ref. \cite{kiehl2023coherence}. All rates were calculated using a cell temperature of $T=108~^{\circ}\text{C}$, buffer-gas pressure of $P_{N_2}=1000$~Torr, and inner cell dimensions of $3\times3\times2$ mm$^3$.}
\end{table}

The simulation of a single FID signal is divided into two steps: pumping and free precession. During the pumping step, we solve Eq.~(\ref{eq:DenMat}) in 100-ns time steps using the equilibrium density matrix $\rho_e$ as the initial state and a nonzero pumping rate $R_{\text{OP}}$ for \SI{10}{\micro\second}. After the pumping step, the pumped density matrix is used as the initial state to solve Eq. (3) with $R_{\text{OP}}=0$. We simulate the free precession in 100-ns time steps for 1 ms. Finally, we compute the expectation value of Faraday rotation for a far-detuned $D_2$ probe beam using the Faraday rotation operator \cite{kiehl2023coherence}
\begin{equation}
\label{eq: FR}
\theta_F = \frac{cr_ef_{D_2}n_\text{Rb}l}{2(2I+1)}\mathcal{L}(\nu-\nu_{D_2})
\begin{cases}
   -m_F &  (F=1)\\
   +m_F & (F=2).
\end{cases}
\end{equation}
Here, $c$ is the speed of light, $r_e$ is the classical electron radius, $f_{D_2}\approx$~2/3 is the $D_2$ oscillator strength, $l=2$ mm is the propagation distance through the atomic media, and $\mathcal{L}(\nu-\nu_{D_2})=(\nu_{D_2}-\nu)/[(\nu-\nu_{D_2})^2+(\Gamma_\text{$D_2$}/2)^2]$ is the value of the line shape with FWHM $\Gamma_\text{$D_2$}$ and optical detuning $\nu-\nu_{D_2}$. For simplicity, we set the constants equal to 1. The Faraday rotation expectation value is $\langle\theta_F\rangle=\text{Tr}(\rho\theta_F)$.
 
\section{\label{sec:FitMethods}Fitting Methods}

There are many ways of extracting the frequency information from an FID signal such as frequency counters \cite{lee2021heading}, demodulation and instantaneous phase retrieval techniques \cite{afach2015highly, wilson2020wide}, and least-squares fitting \cite{hunter2018free, hunter2022accurate, grujic2015sensitive}. Fitting is often desirable for its flexibility, enabling the utilization of different models. Specifically, we focus on single- and double-frequency time-domain fitting models. Other frequency extraction methods that only account for one frequency, such as demodulation, exhibit behavior similar to the single-frequency fitting model discussed here. 

We fit the Faraday rotation signals to the model
\begin{equation}
    V(t)={\sum^{N}_{i=1}}V_i\text{cos}(2{\pi}f_{L,i}t+\phi_i)e^{-t/\tau_i}+V_\text{dc}(t),
\end{equation}
where $V_i$, $f_{L,i}$, $\phi_i$, and $\tau_i$ are the ac amplitude, Larmor frequency, initial phase, and transverse spin-relaxation time for the \textit{i}th frequency component, respectively.  Here, V$_\text{dc}(t)=V_0+V_\text{off}e^{t/T_\text{off}}$ is the dc offset containing an overall offset, $V_0$, and an exponentially decaying offset characterized by an amplitude, $V_\text{off}$, and time constant, $T_\text{off}$. The single-frequency fitting model includes one term in the sum ($N=1$). Given that the separation of the average Zeeman resonance frequencies $f_{L,i}$ in the $F=1$ and $F=2$ manifolds is much larger than the separation between adjacent Zeeman resonance frequencies (1.4 kHz versus 36 Hz at \SI{50}{\micro\tesla}), we can more accurately model low spin-polarization FID signals with a double-frequency fitting model ($N=2$). This model includes a term for each hyperfine manifold. Additionally, we constrain the two frequencies in the double-frequency fitting model to the average Larmor frequency of each hyperfine manifold using the Breit-Rabi formula. This sets the magnetic field as a free parameter of the fit and prevents over-fitting.

\section{\label{sec:PumpDep}Simulated Pump-Power Dependence}

\begin{figure*}[hbt!]
\includegraphics[width=\linewidth]{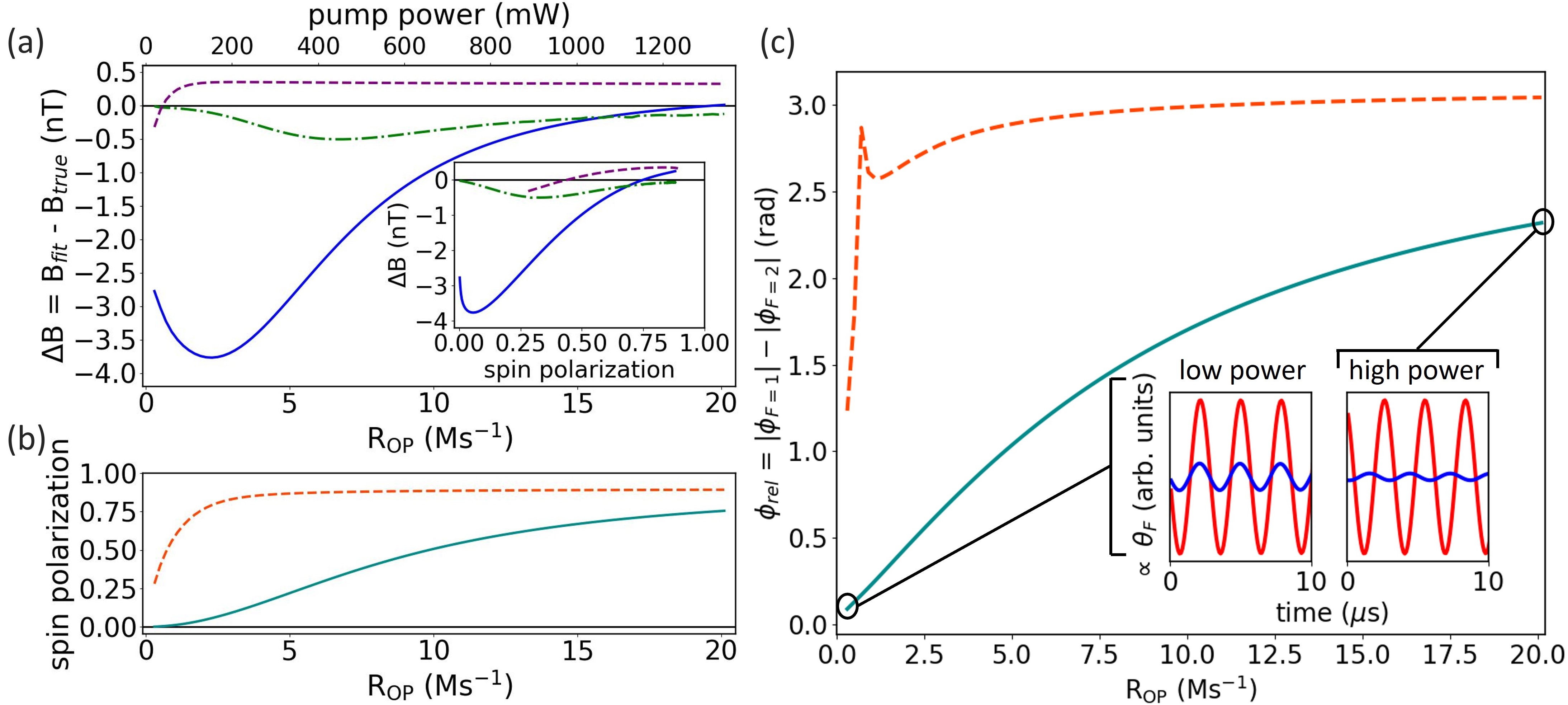}
\caption{\label{fig:SimAcc} (a) Simulated magnetic-field accuracy for SP-SFF (blue solid line), SP-DFF (green dash-dotted line), and SY-SFF (purple dashed line) as a function of pumping rate, $R_{\text{OP}}$ (bottom axis) and pump power (top axis). The pump power was calculated using Eq. (\ref{Eq: Pow}) with a beam waist of 1.5~mm and FWHM of 21.3 GHz. The inset plots the same accuracy data as a function of initial atomic spin polarization $P=2\langle{S_z}\rangle$. (b) Spin polarization versus pumping rate for single-pulse pumping (cyan solid line) and synchronous-pulse pumping (orange dashed line). (c) Simulated relative phase $\phi_{rel}$ versus pumping rate for single-pulse pumping (solid cyan) and synchronous-pulse pumping (dashed orange). The insets show the Faraday rotation $\theta_F$ of the $F=2$ (red) and $F=1$ (blue) manifolds at $R_{\text{OP}}$~=~0.3~Ms$^{-1}$ (left) and $R_{\text{OP}}$~=~20.1~Ms$^{-1}$ (right). The Faraday rotation signals were calculated by splitting the Faraday rotation operator into submatrices for each hyperfine manifold and calculating the expectation value for each hyperfine manifold separately.}
\end{figure*}

To investigate the accuracy of magnetic-field extraction from FID signals at varying spin polarizations, we simulate and analyze FID signals using three methods: single-pulse pumping with single-frequency fitting (SP-SFF), single-pulse pumping with double-frequency fitting (SP-DFF), and synchronous-pulse pumping with single-frequency fitting (SY-SFF). We then compare the extracted-magnetic-field values to the known magnetic-field strength. The obtained spin polarization depends on the mean photon spin \textbf{s}, pumping rate ($R_\text{OP}$), and magnetic field \textbf{B}. Given that a high mean photon spin ($|\textbf{s}|\geq0.9$) can be readily achieved using a quarter wave-plate, we simulate FID signals as a function of pumping rate while keeping the mean photon spin constant. This allows us to compare our numerical simulations with experimental measurements as a function of pumping power because we can relate the pumping rate ($R_\text{OP}$) to the pumping power, $P_\text{pump}$, using the equation \begin{equation}
\label{Eq: Pow}
\begin{aligned}
P_\text{pump}&=\frac{E_\text{ph}A_bR_\text{OP}}{\sigma_\text{abs}}\\
&=~\frac{E_\text{ph}A_\text{b}\Gamma_\text{$D_1$}}{2cr_ef_\text{$D_1$}}R_\text{OP} & (\text{on resonance}),
\end{aligned}
\end{equation} where $E_\text{ph}$ is the energy of the pumping photons, $A_\text{b}$ is the area of the pumping beam written in terms of the beam waist, $\Gamma_\text{$D_1$}$ is the FWHM of the $D_1$ absorption resonance, $\sigma_\text{abs}$ is the absorption cross section of the pumping light, and $f_\text{$D_1$}$ is the $D_1$ oscillator strength \cite{Seltzer2008DevelopmentsIA, hernandez2024cavity}. The second line of this equation holds when the pumping light is resonant with the $D_1$ absorption line.

These simulations assume a magnetic field of \SI{50}{\micro\tesla} in the $\hat{x}$ direction with pumping rates ranging from 0.3 to 20~Ms$^{-1}$ and a mean photon spin $\textbf{s}=0.9\hat{z}$. By orienting a magnetic field perpendicular to the pump-probe laser axis, we minimize inaccuracies from the NLZ effect, allowing the hyperfine systematic error to dominate because the perpendicular field causes the Zeeman resonances to be symmetric, effectively canceling out the NLZ effect. For single-pulse pumping, we use $T_\text{p}=\SI{10}{\micro\second}$ and for synchronous-pulse pumping, we use 200 pulses, each 30 ns long, square-wave modulated between 0 and $R_{\text{OP}}$ at the average $F=2$ Larmor frequency $f_{L, 2}$ (Fig. \ref{fig:FIDBackgorund}(c)). After pumping, we simulate free precession for $T_\text{free}=1$~ms (Fig. \ref{fig:FIDBackgorund}(d)).

\subsection{SP-SFF Method}

The simulated accuracies for each magnetic-field extraction method are shown in Fig.~\ref{fig:SimAcc}(a). The SP-SFF exhibits the largest inaccuracy as the pumping rate decreases. This method exhibits a distinctive pattern consistent with experimental findings (Fig.~\ref{fig:RealData}). At lower pumping rates, the single-frequency fitting method demonstrates a dip in accuracy, with a deviation of approximately 3.5~nT from the true field. As the pumping rate increases, the extracted magnetic field converges toward the true magnetic field.

This systematic error is caused by non-negligible contributions from the $F=1$ state. The separation between the average Larmor frequencies for the $F=1$ and $F=2$ manifolds, $|f_{L,2}-f_{L,1}|$, is proportional to the magnetic field and is approximately 1.4 kHz at \SI{50}{\micro\tesla}. Therefore, coherent Larmor precession of atoms in the $F=1$ manifold shifts the average frequency extracted by a single-frequency fit. As the pumping rate increases, the $F=1$ Larmor signal becomes negligible, making single-frequency extraction techniques accurate approximations. Consequently, the relative amplitude of the $F=1$ Larmor signal is responsible for the saturation of the SP-SFF accuracy in Fig. \ref{fig:SimAcc}(a). 

However, the relative amplitude alone does not explain the behavior of the SP-SFF accuracy at low pumping rates. A complete understanding necessitates considering the relative phase $\phi_{rel}$ between the $F=1$ and $F=2$ Larmor precession signals. Because the atoms in the $F=1$ and $F=2$ manifolds precess in opposite directions, it is commonly assumed that the relative phase is $\pi$. Using an analytical model described in Appendix \ref{App: AM}, we can extract the relative phase from the density matrices used in the simulations of Fig.~\ref{fig:SimAcc}(a). The phases for each Larmor transition within a manifold are approximately equal. However, the relative phase between different manifolds varies depending on the initial density matrix after pumping, which is influenced by the pumping rate and method (Fig. \ref{fig:SimAcc}(c)). This relative phase arises from the steady-state orientation of the average $F=I~{\pm}~1/2$ atomic spins after optical pumping, a concept further explored in Sec. \ref{sec:PhaseShifts}. As the pumping rate goes to zero, $\phi_{rel}\rightarrow0$ and as the pumping rates goes to infinity, $\phi_{rel}\rightarrow\pi$. 

The insets of Fig. \ref{fig:SimAcc}(c) depict the average $F=1$ (blue) and $F=2$ (red) Faraday rotation signals after optical pumping. These Faraday rotation signals were calculated by splitting the Faraday rotation operator into $3\times3$ and $5\times5$ sub-matrices for each hyperfine manifold and computing their expectation values separately. From these insets, we see that at low pumping rates ($R_\text{OP}$~=~0.3~Ms$^{-1}$), the $F=I~{\pm}~1/2$ Faraday rotation signals are approximately in phase. Conversely, at high pumping rates ($R_\text{OP}$~=~20~Ms$^{-1}$), these signals approach a phase difference of nearly $0.8\pi$.

The shifting relative phase alters the interference between the average $F=1$ and $F=2$ Larmor precession signals, thereby affecting the accuracy of single-frequency extraction. Further detailed in Appendix \ref{App: Amp Phase}, this varying relative phase is responsible for the observed dip in the SP-SFF accuracy at low pumping rates (Fig. \ref{fig:SimAcc}(a)). It is worth mentioning that two orthogonal probe beams can observe this relative phase and thereby distinguish between the $F=1$ and $F=2$ signals \cite{lee2021heading}.

\begin{figure*}[hbt!]
\includegraphics[width=\linewidth]{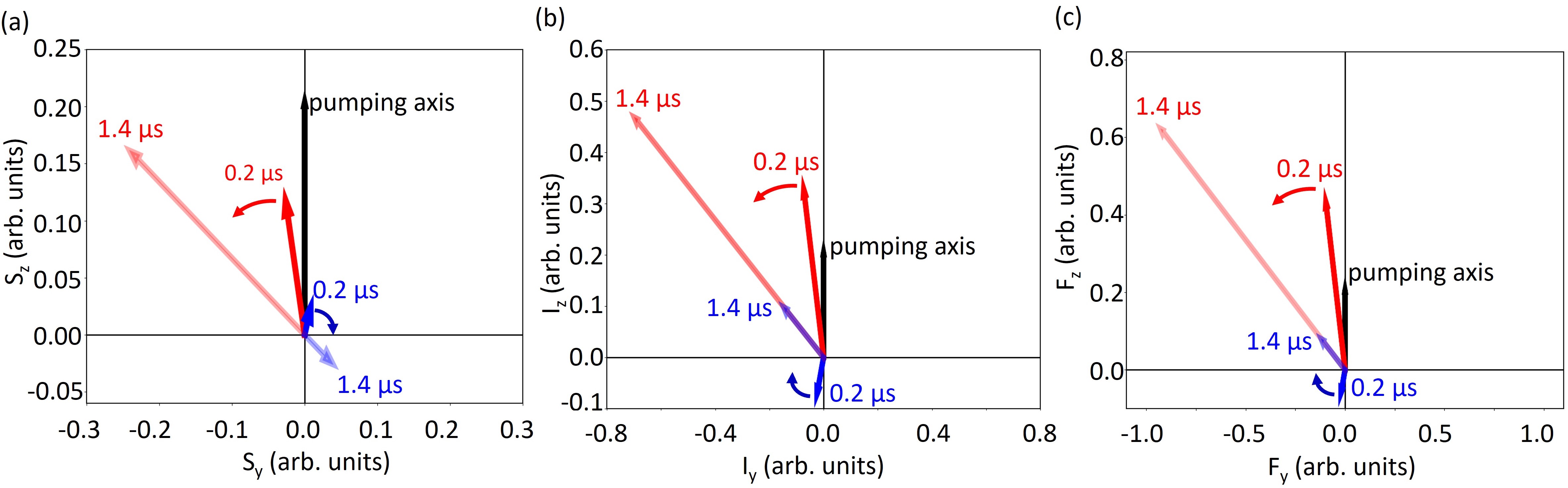}
\caption{\label{fig:AllSpinsvsTime} Expectation values for (a) electron, (b) nuclear, and (c) total atomic spins during optical pumping. The red (blue) vectors represent the $F=2$ ($F=1$) spin component. This simulation was performed with $R_{\text{OP}}$~=~5~Ms$^{-1}$ at a \SI{50}{\micro\tesla} magnetic field. The black vector indicates the laser axis. The horizontal axis represents the spin component along the $y$ axis, while the vertical axis represents the spin component along the $z$ axis. The darker vector represents the spin after \SI{0.2}{\micro\second} of optical pumping and the lighter vector represents the spin after \SI{1.4}{\micro\second} of optical pumping.}
\end{figure*}

\subsection{SP-DFF Method}

In regimes of low spin polarization, i.e. low pumping power and/or small mean photon spin, the systematic inaccuracy associated with the relative amplitude and phase between the average $F=1$ and $F=2$ precession can be mitigated by using the double-frequency fitting model described in Sec. \ref{sec:FitMethods}. The accuracy associated with SP-DFF is shown by the green dash-dotted line in Fig. \ref{fig:SimAcc}(a). The simulated accuracy remains under about 0.5~nT within the investigated pumping rates. Comparing the worst accuracies, the SP-DFF method has an accuracy enhancement of roughly 7 (3.5~nT/0.5~nT) compared to the SP-SFF method. In contrast to single-frequency fitting, the double-frequency fitting model provides a more accurate estimation of the magnetic field at low spin polarizations by accounting for the relative amplitude and phase of both $F=1$ and $F=2$ Larmor precession signals. Since the frequency separation between adjacent Zeeman transitions (36 Hz) is much less than the frequency separation between the average Zeeman transitions for the two hyperfine manifolds (1.4 kHz), extending the model to include all six Larmor frequencies present in a $^{87}$Rb FID signal does not significantly enhance the accuracy.  

\subsection{SY-SFF Method}

The hyperfine systematic error can also be mitigated using synchronous-pulse pumping. The purple dashed line in Fig. \ref{fig:SimAcc}(a) displays the accuracy associated with SY-SFF. Since synchronous-pulse pumping is a resonant process, high spin polarizations can be achieved with relatively low pumping rates compared to single-pulse pumping (Fig.~\ref{fig:SimAcc}(b)). Additionally, the relative phase, $\phi_{rel}$, approaches $\pi$ at lower pumping rates compared to single-pulse pumping (Fig. \ref{fig:SimAcc}(c)). This resonant process minimizes the $F=1$ contribution to the Larmor precession signal, making single-frequency extraction methods more accurate. The simulated accuracy using SY-SFF is under 0.5 nT for the pumping rates investigated here. Similar to SP-DFF, this method also has an accuracy enhancement of roughly 7 (3.5~nT/0.5~nT) relative to SP-SFF when comparing the most inaccurate data points. The peak of $\phi_{rel}$ at 0.7~Ms$^{-1}$ in Fig. \ref{fig:SimAcc}(c) is not well understood and warrants further investigation.

Despite quickly saturating to the highest possible spin polarization of $P=0.9$ (limited here by the mean photon spin $|\textbf{s}|=0.9$), the magnetic field extracted using the SY-SFF technique has a bias of approximately 0.3 nT. Similarly, when the SP-SFF technique also reaches the maximum possible spin polarization $P=0.9$, the extracted field error is approximately 0.3 nT. This is due to imperfect spin polarization $P~\textless~1$ and inaccuracies in the average gyromagnetic ratio. If we use a perfect spin polarization of $P=1$, then the inaccuracies due to the hyperfine systematic error are suppressed to the picotesla level. In this work, we define the gyromagnetic ratio as $\gamma$ = $f_{L,2}$/B~=~6.99579~Hz/nT, where $f_{L,2}$ is the average $F=2$ Larmor frequency. Furthermore, the inset of Fig. \ref{fig:SimAcc}(a) shows that the accuracy at the same spin polarization is different for the SY-SFF and SP-SFF techniques. This occurs because different initial distributions can result in the same initial atomic spin polarization, and these distributions will have nonidentical amplitudes for each Larmor frequency component, leading to differences in accuracy.

Both the SP-DFF and the SY-SFF techniques  successfully mitigate the $F=1$ systematic error, with their respective accuracies consistently agreeing within 1 nT of each other throughout all the explored pumping rates. The advantage of the SP-DFF method is that it does not require prior knowledge of the magnetic field, whereas the SY-SFF method necessitates either prior field information or the implementation of a feedback loop to maintain resonance between the synchronous pulses and the Larmor precession. The SY-SFF method's advantage lies in its highly efficient pumping which enhances the accuracy of single-frequency extraction techniques such as demodulation or frequency counting. Although not explored in this study, `enhanced optical pumping' (EOP) can generate high spin polarizations with relatively low optical power by applying a strong bias field (on the order of millitesla) along the laser axis during optical pumping \cite{hunter2023optical}. Like the SY-SFF method, EOP also reduces the hyperfine systematic error since the $F=1$ contributions are negligible. 

It is important to note that these methods do not mitigate NLZ shifts within a hyperfine manifold. Therefore, they should be paired with another technique, such as the analytical correction outlined in \cite{lee2021heading}, to improve magnetometer accuracy across all field directions and spin polarizations.

\section{\label{sec:PhaseShifts}Relative Phase Effect}

Grasping the factors that influence the relative phase is crucial, as it affects the accuracy of an FID magnetometer. The relative phase results from the initial orientation of the average $F=I~{\pm}~1/2$ spins after pumping. Therefore, understanding the relative phase necessitates understanding optical pumping. In this section we focus on examining the relative phase after single-pulse pumping and then extrapolate these insights to the case of synchronous-pulse pumping.

Since the \SI{50}{\micro\tesla} magnetic field is oriented in the $\hat{x}$ direction, the spins precess in the $yz$ plane (Fig.~\ref{fig:FIDBackgorund}). As before, we use a mean photon spin of \textbf{s}~=~0.9$\hat{z}$, meaning that the pumping axis is oriented in the $\hat{z}$ direction resulting in $\sigma^+$ transitions. We calculate the electron, nuclear, and total atomic spins in the $|F,~m_F\rangle_z$ basis for both the $F=1$ and $F=2$ manifolds by splitting the spin operators into $3\times3$ and $5\times5$ submatrices and calculating their expectation values. Figures~\ref{fig:AllSpinsvsTime}(a), \ref{fig:AllSpinsvsTime}(b), and \ref{fig:AllSpinsvsTime}(c) depict the electron, nuclear, and total atomic spins, respectively, during single-pulse pumping.

Due to opposite signs of the $F=1$ and $F=2$ $g$-factors, the respective spins rotate in opposite directions. Furthermore, in the $F=2$ manifold, the average electron and nuclear spins are parallel, while in the $F=1$ manifold, they are antiparallel \cite{Seltzer2008DevelopmentsIA}. The pumping light interacts strongly with the valence electron but not directly with the nucleus, because it is shielded by inner electron shells. However, the nucleus indirectly interacts with the pumping light via the hyperfine coupling with the electron. 

Since the electron strongly interacts with the pumping light, the electron spin is initially pumped parallel to the pumping axis (+$\hat{z}$) for both hyperfine manifolds, as observed at \SI{0.2}{\micro\second} after the pumping light is switched on in Fig. \ref{fig:AllSpinsvsTime}(a). In the $F=2$ manifold, the nuclear spin is also initially pumped parallel to the pumping axis because it must be parallel to the electron spin. Conversely, in the $F=1$ manifold, the nuclear spin is initially pumped antiparallel to the pumping axis because it must be antiparallel to the electron spin [\SI{0.2}{\micro\second} vector in Fig. \ref{fig:AllSpinsvsTime}(b)]. Because the nucleus stores most of the angular momentum ($I=3/2$ versus $S=1/2$), the average total atomic spin mirrors the behavior of the average nuclear spin. Consequently, the $F=1$ total atomic spin is initially pumped antiparallel to the pumping axis, polarizing the $F=1$ manifold towards the $|1, -1\rangle_z$ state (Fig.~\ref{fig:AllSpinsvsTime}(c)).

Eventually, the atoms reach a steady state between the precession caused by the magnetic field and the optical pumping. For the 5~Ms$^{-1}$ pumping rate used in these simulations, the spins reach their steady state after approximately \SI{1.4}{\micro\second} of pumping. Interestingly, repopulation pumping causes the average $F=1$ spin to rotate more than the average $F=2$ spin (Fig.~\ref{fig:AllSpinsvsTime}). Initially pumped antiparallel to the pumping axis, into the $|1, -1\rangle_z$ state, repopulation rotates the average $F=1$ spin towards the pumping axis, moving population into the $|1, 1\rangle_z$ state.

\begin{figure}[hbt!]
\includegraphics[width=\linewidth]{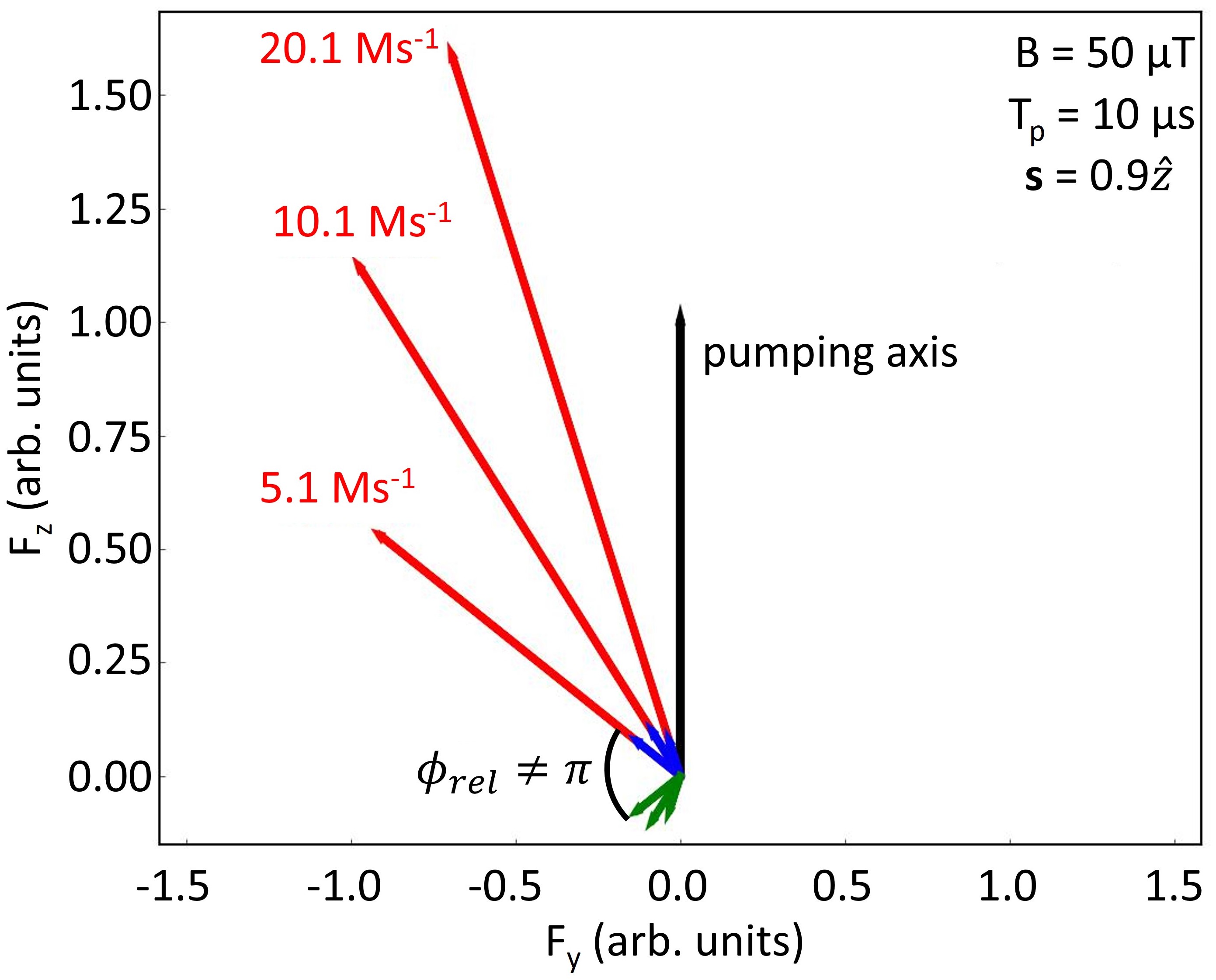}
\caption{\label{fig:AtomicSpinsvsRp} Simulated total atomic spins after single-pulse pumping for \SI{10}{\micro\second} with $\sigma^+$ photons at various pumping rates with a \SI{50}{\micro\tesla} field oriented in the $\hat{x}$ direction. The blue vectors are the total atomic spin for the $F=1$ manifold and the red vectors are the total atomic spin for the $F=2$ manifold. The black arrow represents the pumping laser axis. The green vectors indicate the $F=1$ atomic spin as measured by the probe laser. Here the $z$ component of the $F=1$ atomic spin is multiplied by $-1$ to account for the Faraday rotation as described by Eq.~(\ref{eq: FR}). The angle between the $F=2$ vector (red) and the Faraday rotated $F=1$ vector (green) represents the relative phase, $\phi_{rel}$.}
\end{figure}

\begin{figure*}[hbt!]
\includegraphics[width=\linewidth]{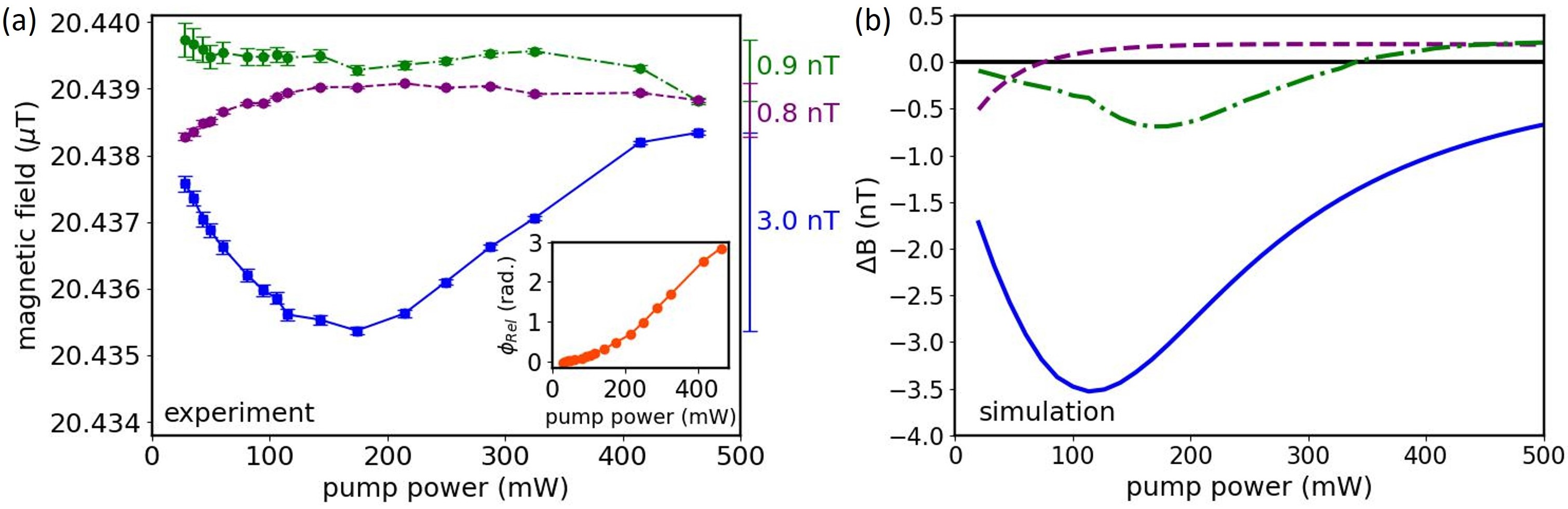}
\caption{\label{fig:RealData} (a) Experimentally measured pump-power dependence of the extracted magnetic field. The signals generated via single-pulse pumping data are fit to the single-frequency model (blue solid line with squares) and the double-frequency model (green dash-dotted line with circles). The synchronously pumped data are fit to the single-frequency model (purple dashed line with circles). The error bars are the standard deviations of the magnetic field outputted from the fits. The inset of this graph plots the average relative phase between the $F=2$ and $F=1$ manifolds extracted from the double-frequency fits. (b) Simulated magnetic-field accuracy at a bias field of \SI{20.435}{\micro\tesla}. The blue solid line is the SP-SFF accuracy, the green dash-dotted line is the SP-DFF accuracy, and the purple dashed line is the SY-SFF accuracy.}
\end{figure*}

In the presence of buffer gas, collisional mixing quickly depolarizes electron spins in the excited state, leading to an equal probability of decaying into either ground-state hyperfine manifold \cite{Seltzer2008DevelopmentsIA}. However, these collisions are sudden with respect to the nuclear spins such that the atoms preserve their nuclear spin upon deexcitation \cite{appelt1998theory}. Therefore, the atoms do not have equal probability of decaying into each hyperfine sublevel but are restricted to the sublevels that preserve the nuclear angular momentum. After an interval of optical pumping, a majority of the population lies in the $F=2$ manifold, biased towards the positive $m_F$ sublevels which carry positive nuclear angular momentum. After these atoms are excited, half of them decay into the $F=1$ manifold, mainly repopulating the $|1, 1\rangle_z$ and $|1, 0\rangle_z$ states, flipping the direction of the average $F=1$ spin. Combined with the precession caused by the magnetic field, the repopulation effect causes the $F=1$ spin vectors to rotate more during optical pumping than the $F=2$ spin vectors such that the $F=1$ and $F=2$ spins reach a steady state nearly parallel to each other (Fig.~\ref{fig:AllSpinsvsTime}(c)). 

As shown in Fig.~\ref{fig:AtomicSpinsvsRp}, a higher pumping rate results in the steady-state total atomic spins approaching closer to alignment with the pumping axis. The probe laser, aligned with the $\hat{z}$ direction, exclusively detects the $z$ component of atomic spins. Since the Faraday rotation operator is proportional to $-m_F$ in the $F=1$ manifold [Eq.~(\ref{eq: FR})], we illustrate the effect of Faraday rotation by multiplying the $z$ component of the $F=1$ atomic spin by $-1$ (green vector). The angle between the $F=2$ atomic spin vector (red) and the $F=1$ atomic spin vector as seen by the probe laser (green) matches the relative phase extracted from the density matrix.

In the infinite pumping rate limit, the steady-state spins perfectly align with the pumping axis (Fig. \ref{fig:AtomicSpinsvsRp}). Consequently, the Faraday rotation has a phase shift of $\pi$ between the $F=1$ and $F=2$ Larmor signals. Conversely, at very low pumping rates, the spins take longer to reach a steady state, potentially exceeding the duration of the pumping pulse. Given the comparable magnitude of the $F=1$ and $F=2$ $g$-factors, the spins will oscillate in opposite directions with the average $F=2$ spin starting parallel and the average $F=1$ spin starting antiparallel to the pumping axis such that the relative phase of the respective Larmor signals is always approximately zero. In between these limits, the average $F=1$ and $F=2$ spins reach a steady state such that the relative phase lies within 0~${\leq}~{\phi_{rel}}~{\leq}~\pi$.

The physics is the same for synchronous-pulse pumping schemes. The main difference is that the spins are allowed to precess an integer number of Larmor periods between pump pulses. Since synchronous-pulse pumping is more efficient than single-pulse pumping, the initial atomic spins are oriented closer to the pumping axis \cite{hunter2018free, gerginov2020scalar}. Therefore, the relative phase converges to $\pi$ at much lower pumping rates compared to single-pulse pumping (Fig. \ref{fig:SimAcc}(c)).

\section{\label{sec:RealData}Fitting Experimental Data}

To confirm the theory laid out in the previous sections, we measured FID signals at various optical pumping powers from 28 to 465 mW while keeping the probe power fixed at 5 mW. The pumping laser frequency was resonant with the $D_1$ transition, while the probe laser frequency was approximately $-80$ GHz detuned from the $D_2$ transition. A simplified diagram of our experimental setup is shown in Fig. \ref{fig:FIDBackgorund}(b). The measurements were conducted in a magnetic field of approximately \SI{20.4}{\micro\tesla}, oriented in the $\hat{x}$ direction. We measured 100 FID signals at each pumping power for both single-pulse pumping and synchronous-pulse pumping. In the single-pulse pumping scheme, we used $T_\text{p}=\SI{10}{\micro\second}$, while the synchronous-pulse pumping scheme used 200 30-ns pulses modulated at the average Larmor frequency.

Figure~\ref{fig:RealData}(a) displays the mean magnetic field extracted from the FID signals across a range of pumping powers using the SP-SFF, SP-DFF, and SY-SFF methods. Figure~\ref{fig:RealData}(b) displays the simulated accuracy with the same parameters as before but at approximately the same magnetic field as the experimental data. The extracted magnetic field associated with the SP-SFF method (blue solid line) has the same distinctive pattern observed in the simulated data of Fig.~\ref{fig:RealData}(b). The simulations revealed that the dip in the extracted field at low pumping powers can be attributed to the relative phase between the $F=1$ and $F=2$ Larmor precession signals, while the saturation of the extracted magnetic field at high pumping powers can be attributed to the diminishing amplitude of the $F=1$ signal. Indicated by the blue bar on the right side of Fig. \ref{fig:RealData}(a), the extracted magnetic field using SP-SFF spans approximately 3 nT, aligning closely with our simulation.

By fitting the same single-pulse pumped data with the double-frequency fitting model, we can minimize the systematic error caused by the $F=1$ contributions to the FID signal, as demonstrated by the green dash-dotted line in Fig.~\ref{fig:RealData}. Utilizing the SP-DFF technique, the extracted magnetic field demonstrates a diminished dependence on pumping power, spanning about 0.9 nT, highlighted by the green bar on the graph's right side. Comparing the ranges of the extracted field, this is an enhancement of approximately 3.3 (3 nT/0.9 nT) compared to the SP-SFF method. Just as in the simulation, this is attributed to the double-frequency fitting model accounting for the relative amplitude and phase of the $F=1$ Larmor signal. Moreover, the relative phase extracted from the fit parameters of the double-frequency fits, depicted in the inset of Fig. \ref{fig:RealData}, mirrors the relative phase behavior observed in the simulations shown in Fig. \ref{fig:SimAcc}. At low pumping powers, the relative phase approaches zero and as the pumping power increases the relative phase increases to $\pi$. 

The purple dashed line in Fig.~\ref{fig:RealData} shows the extracted magnetic field using the SY-SFF method. Synchronous-pulse pumping efficiently drives the atomic ensemble into the stretched state, minimizing $F=1$ contributions even at lower pumping rates (Fig.~\ref{fig:SimAcc}(b)). This reduces the dependence on pumping power. The magnetic field extracted using the SY-SFF technique spans 0.8 nT, as indicated by the purple bar on the right side of the graph, which is an enhancement of 3.75 (3~nT/0.8~nT) relative to the SP-SFF method.

Significantly, both SP-DFF and SY-SFF magnetic-field-extraction techniques exhibit agreement, demonstrating discrepancies of less than 1.5 nT between each other at all investigated pumping powers. This agreement is apparent when inspecting the overlapping range bars for both data sets. This, coupled with the outcomes of our simulations, strongly suggests that the magnetic field extracted through the SP-DFF and SY-SFF techniques offers superior accuracy across an expanded range of spin polarizations compared to the magnetic field extracted utilizing the SP-SFF technique. Notably, the extracted magnetic fields from both the SP-DFF and SY-SFF techniques deviate from each other at the lowest pumping powers. This might be explained by a decreasing signal-to-noise ratio or by field drifts between measurements which were not included in the simulations. While we cannot assert absolute accuracy for the magnetometer as there may be other systematic effects like light shifts, nonlinear Zeeman shifts, and magnetic-field gradients, these improved magnetic-field-extraction techniques have effectively mitigated inaccuracies caused by the hyperfine systematic error to the $\pm1$~nT level.

\section{\label{sec:Concl}Conclusion}

We studied the accuracy of extracting the magnetic field from FID signals over a wide range of spin polarizations. As the spin polarization changes, the relative amplitudes and phases between the $F=1$ and $F=2$ Zeeman frequency components change, ultimately affecting the accuracy of magnetic-field extraction. Simulations reveal that $F=1$ contributions introduce inaccuracies of approximately 3.5 nT when employing the SP-SFF method. Theoretically, these inaccuracies can be reduced to under 0.5 nT through the utilization of either SP-DFF or SY-SFF techniques. Measurements of the extracted magnetic field closely match the simulations, displaying the same characteristic patterns. The variation of the magnetic field extracted from the SP-SFF measurements is approximately 3 nT, while the variations of the magnetic field extracted from the SP-DFF and SY-SFF methods are both under 1 nT. Furthermore, the magnetic fields measured using the SP-DFF and SY-SFF techniques agree to within 1.5~nT at all pumping powers investigated. 

The SP-DFF technique is well suited for accurately extracting the magnetic field from FID signals in the low-spin-polarization regime $P\leq0.5$. Furthermore, it does not require feedback loops or prior knowledge of the magnetic field to achieve high accuracies. In contrast, the SY-SFF scheme generates higher spin polarizations at lower pumping rates, making single-frequency extraction techniques such as demodulation, instantaneous-phase retrieval, or frequency counters more accurate. However, this enhancement necessitates prior knowledge of the magnetic field to guess the pulse-modulation frequency. While the techniques introduced in this paper do not directly address absolute accuracy or the NLZ effect, they effectively mitigate inaccuracies linked to the overlap of $F=1$ and $F=2$ Zeeman resonances. Compact sensors utilizing microfabricated cells with high buffer-gas pressure, operating at elevated temperatures and technically constrained by moderate optical power levels, are particularly susceptible to this systematic inaccuracy. The techniques described in this paper can significantly improve the accuracy of such sensors under limited optical power conditions.

\begin{acknowledgments}
We wish to acknowledge helpful discussions with Ricardo Jiménez-Martínez, Georg Bison, and Tobias Thiele. This work was supported by the DARPA SAVaNT program through grant W911NF-21-1-0127, and NSF QLCI Award OMA-2016244.
\end{acknowledgments}

\appendix

\section{Analytical Model for Extracting Phases}
\label{App: AM}

\begin{figure}[hbt!]
\includegraphics[width=\linewidth]{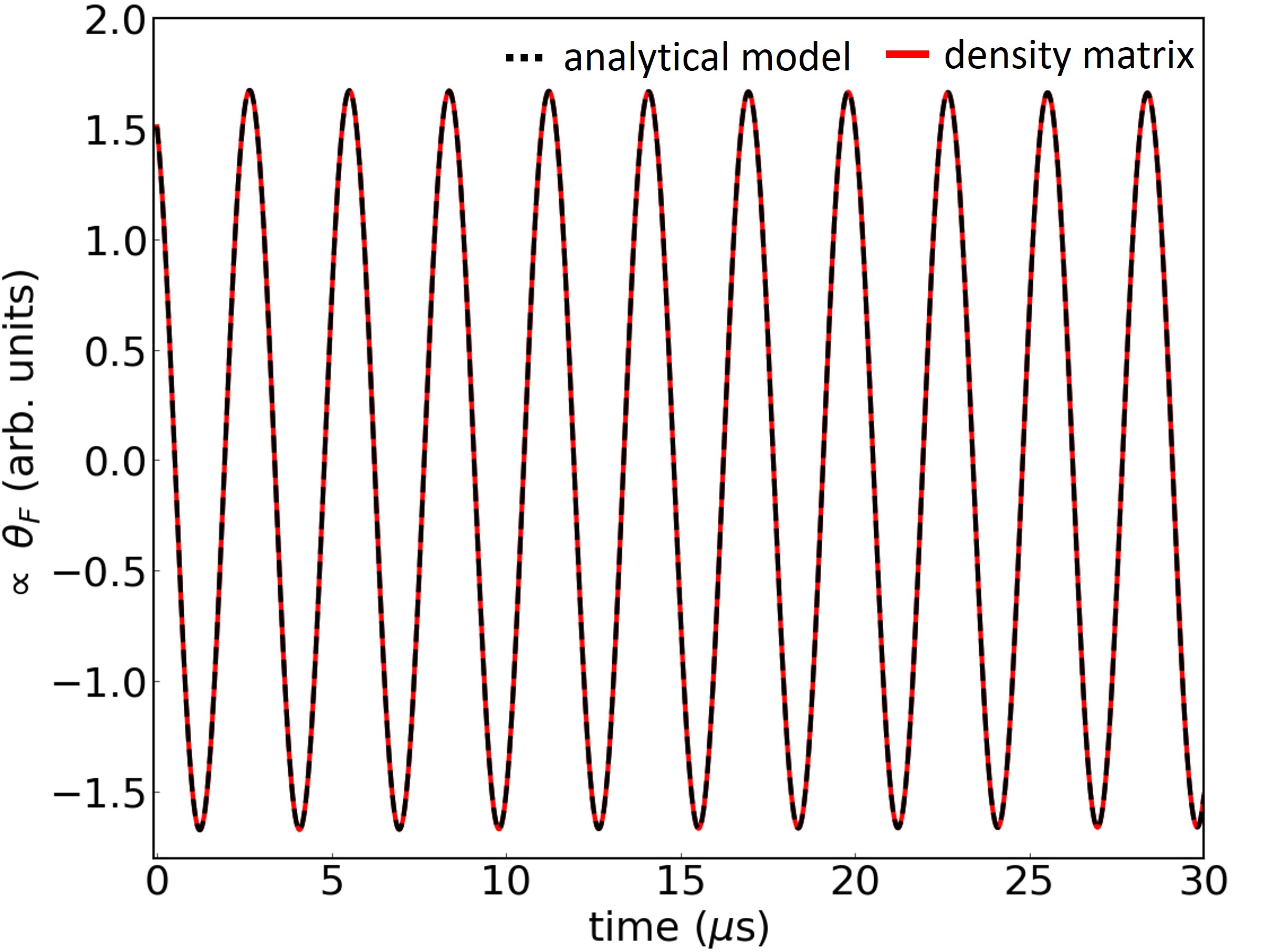}
\caption{\label{fig:AnalyticalMod} Faraday rotation signals without decoherence simulated using the full density-matrix model (red solid line) and the analytical model (black dashed line). The consistency between the analytical model and the density matrix highlights the analytical model's ability to accurately predict the phases of each Larmor transition.}
\end{figure}

We can write an analytical model that does not include decoherence to help us understand the physics of the system and approximate parameters such as the phase of each Larmor transition. We start by diagonalizing the Hamiltonian from Eq. (4), $\tilde{H}{\rightarrow}\tilde{H}_d$. Next we convert the Faraday rotation operator and the input density matrix into the basis that diagonalizes the Hamiltonian, $\theta_F{\rightarrow}\theta_{F,d}=U{^\dag}{\theta_F}U$ and $\rho_0{\rightarrow}\rho_{0,d}=U{^\dag}{\rho_0}U$, where $U$ is the unitary matrix of eigenvectors that diagonalizes the Hamiltonian. The Schr\"{o}dinger equation can be solved without decoherence by $\rho_{0,d}$(t) = $e^{-iH_dt/{\hbar}}{\rho_{0,d}}^{iH_dt/{\hbar}}$. The Faraday rotation can be calculated using $\theta_F~{\propto}~\text{Tr}[\theta_{F,d}\rho_{0,d}(t)]$. Finally, we find that the Faraday rotation can be calculated at any time based on the input density matrix using the equation 
\begin{equation}
    \begin{aligned}
        \theta_F \propto &\sum_{i=1}^{8} F_d^{ii}{\rho_{0,d}^{ii}} \\
    & + \sum_{i{\neq}j=1}^{8} 2|F_d^{ij}{\rho_{0,d}^{ji}}|\text{cos}[2{\pi}(f_i-f_j)t+\text{arg}(F_d^{ij}{\rho_{0,d}^{ji}})].
    \end{aligned}
\end{equation}
The first term accounts for the dc offset, while the second term represents the precession. The argument of the cosine consists of two terms: The first term is the frequency and the second term is the phase. Figure \ref{fig:AnalyticalMod} shows the excellent agreement between the analytical model and density-matrix simulations. For this density-matrix simulation, we included decoherence during pumping, but not during free precession.

The agreement observed here demonstrates the analytical model's ability to accurately predict the initial phases of the simulated FID signals, even when decoherence effects are included during pumping. Furthermore, the phase of each transition, calculated using arg($F_d^{ij}{\rho_{0,d}^{ji}}$), are approximately equal within a hyperfine manifold. Consequently, we can precisely determine relative phases by computing the difference between the absolute values of the average $F=1$ and $F=2$ phases.

\section{Separating the Effects of Relative Amplitudes and Phases on the Accuracy}
\label{App: Amp Phase}

\begin{figure}[hbt!]
\includegraphics[width=\linewidth]{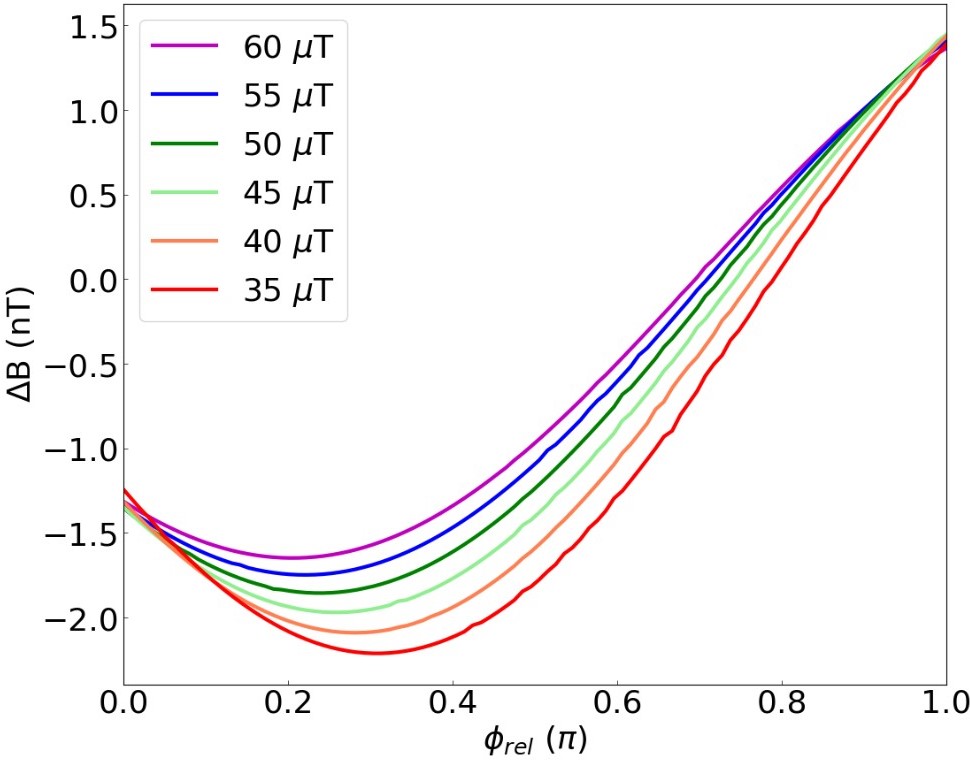}
\caption{\label{fig:ToyRelPhase} Accuracy of single-frequency magnetic field extraction versus relative phase, $\phi_{rel}$, at various magnetic fields. The $F=2$ and $F=1$ amplitudes and time constants are fixed while the $F=2$ and $F=1$ phases are varied from $\pi/2$ to 0 and $\pi/2$ to $\pi$, respectively, in 100 evenly spaced steps.}
\end{figure}

To show that the interference caused by the relative phase between the hyperfine manifolds contributes to the fitting inaccuracies, we fit fake signals generated using the double-frequency fitting equation to the single-frequency fitting model. Using the Breit-Rabi formula, the magnetic field constrains the frequencies of each manifold. The dc offset is set to 0. To approximate an FID signal, the $F=2$ and $F=1$ amplitudes are fixed to 1.5 and 0.25, respectively, while the $F=2$ and $F=1$ time constants are fixed to 1 and 0.25 ms, respectively. We vary the $F=2$ phase from $\pi/2$ to 0 and the $F=1$ phase from $\pi/2$ to $\pi$ in 100 evenly spaced steps. Figure \ref{fig:ToyRelPhase} displays the magnetic-field-extraction accuracy using the single-frequency fitting equation for various magnetic fields. This graph shows that the accuracy of fitting a signal made up of two frequencies varies with the relative phase.

\begin{figure}[H]
\includegraphics[width=\linewidth]{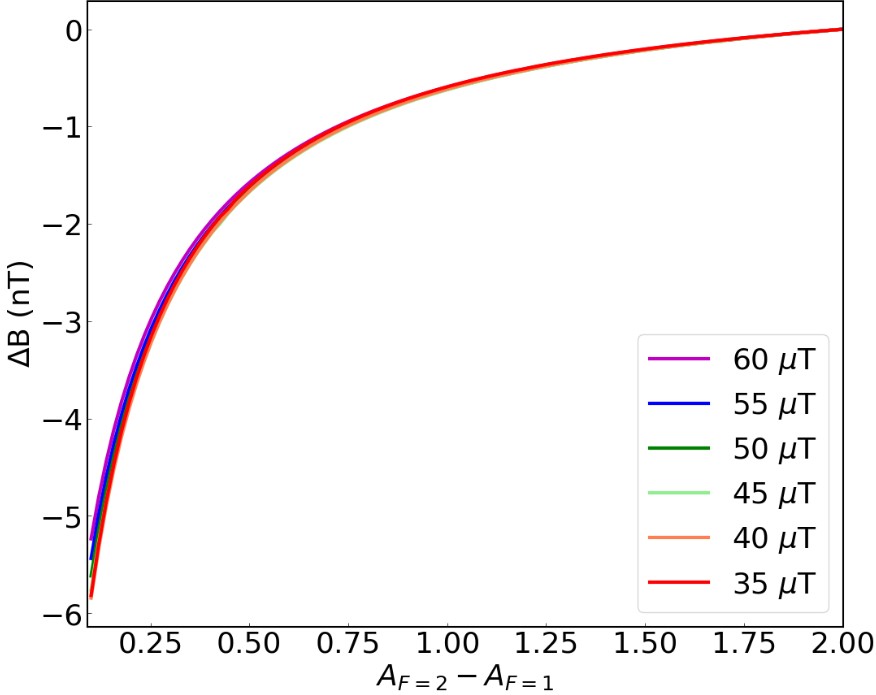}
\caption{\label{fig:ToyAmps} Magnetic-field accuracy of the single-frequency fit as the amplitudes are varied. The horizontal axis of this plot is the $F=2$ amplitude $A_{F=2}$ minus the $F=1$ amplitude $A_{F=1}$. Here the time constants and phases are fixed, while the $F=2$ and $F=1$ amplitudes are varied from 0.25 to 2 and from 0.15 to 0, respectively, in 100 evenly spaced steps. }
\end{figure}

Since we keep the amplitudes fixed in the rudimentary simulation of Fig. \ref{fig:ToyRelPhase}, we do not see the saturation of the single-frequency magnetic field accuracy in the high spin-polarization limit as described in  Sec. \ref{sec:PumpDep}. We can separate the saturation from the interference by fixing the $F=2$ and $F=1$ time constants to 1 and 0.25 ms and the phases to 0 and $\pi$, respectively, while varying the $F=2$ amplitude $A_{F=2}$ from 0.25 to 2, and the $F=1$ amplitude $A_{F=1}$ from 0.15 to 0 in 100 evenly spaced steps. We see from Fig. \ref{fig:ToyAmps} that as the $F=2$ population starts to dominate the signal, the magnetic field extracted from the single-frequency fit saturates to the true value. These two rudimentary simulations confirm that the accuracy saturation is linked to the relative amplitudes while the dip is linked to the relative phase between hyperfine manifolds.

\bibliography{FittingPaper.bib}

\end{document}